# Absence of local moments in the kagome metal $KV_3Sb_5$ as determined by muon spin spectroscopy

Eric M. Kenney[1], Brenden R. Ortiz[2], Chennan Wang[3], Stephen D. Wilson[2], and Michael J. Graf[1]

[1] Department of Physics, Boston College, Chestnut Hill, MA 02467, United States
[2] Materials Department, University of California Santa Barbara, Santa Barbara, CA 93106-9010, United States
[3] Laboratory for Muon Spin Spectroscopy, Paul Scherrer Institute, 5232 Villigen, Switzerland

We have carried out muon spin relaxation and rotation measurements on the newly discovered kagome metal $KV_3Sb_5$, and find a local field dominated by weak magnetic disorder which we associate with the nuclear moments present, and a modest temperature dependence which tracks the bulk magnetic susceptibility. We find no evidence for the existence of $V^{4+}$ local moments, suggesting that the physics underlying the recently reported giant unconventional anomalous Hall effect in this material warrants further studies.

The triangle-based kagome crystalline structure hosts a wide variety of interesting phenomena. Most attention has focused on insulating materials for which local moments occupy the kagome sites, introducing geometric frustration of magnetic ordering and the possibility of forming quantum spin liquids and similar novel magnetic ground states [1-3]. Less well-studied are conducting kagome materials [4-6]. Recently, conducting Kagome materials $AV_3Sb_5$, with $A$ = K, Rb, and Cs, have been synthesized and characterized [7]. The vanadium sublattice within the V-Sb layers forms a two-dimensional kagome net, with potassium intercalated between the V-Sb layers. DFT calculations indicate that the these materials are conducting and have Dirac bands near to the Fermi surface, suggesting that topological effects may play an important role in the observed properties; the Dirac bands have been observed by ARPES measurements in $KV_3Sb_5$ [8]. $CsV_3Sb_5$ has recently been discovered to be a superconductor with $T_c$ = 2.5 K [9]. The predicted existence of several protected Dirac crossings in the $Z_2$ topological metal is suggestive of unconventional superconductivity.

The nature of magnetism in these materials has not been resolved. The bulk magnetic susceptibility for all the $AV_3Sb_5$ materials is relatively small, which likely arises from a high degree of covalency in this compound. Nevertheless, in an ionic limit when using simple electron counting one magnetic $V^{4+}$ ion is predicted per formula unit. The temperature dependent magnetization of $KV_3Sb_5$ exhibits an anomaly at 80 K, but elastic neutron scattering shows no evidence for either a structural transition or magnetic ordering [7]. Recent magnetotransport measurements on $KV_3Sb_5$ showed a large and unconventional anomalous Hall effect (UAHE) [8], which was attributed to skew scattering of topological carriers by spin clusters of three local moments, as studied theoretically in Ref. 10. The local magnetic probe technique of positive muon spin relaxation/rotation ($\mu^+$SR) is an excellent tool to study magnetism and possible spin cluster formation as it is extremely sensitive to both very small local fields and the presence of magnetic inhomogeneity [11].

In this work we report $\mu^+$SR measurements on $KV_3Sb_5$ at temperatures down to 1.6 K. Surprisingly, we find no evidence for local $V^{4+}$ moments, and that the muon depolarization is dominated by the much weaker nuclear moments present, with a slight temperature dependence mirroring that of the bulk susceptibility. These results suggest that mechanisms other than scattering from spin clusters should be investigated in order to explain the observed UAHE.

$KV_3Sb_5$ powder was synthesized via mechanochemical methods from elemental reagents K (Alfa, 99.8%), V (Sigma, 99.9%), and Sb (Alfa, 99.999%). A small excess of K was used (corresponding to $K_{1.05}V_3Sb_5$) to compensate for volatility of the alkali metal. Reagents were loaded into tungsten carbide ball-mill vials and ground for 1h, and subsequently annealed for 48h at 600C in evacuated fused silica ampoules. The powder quality was evaluated through X-ray diffraction, and confirming the phase purity of the $KV_3Sb_5$ powders with excellent crystallinity.

The resultant material was pressed into a disk of 10 mm diameter and 3 mm thickness. The sample was then studied via the local magnetic probe technique of positive muon spin spectroscopy ($\mu^+$SR) [11]. Experiments were performed on the piM3 beamline using a continuous flow cryostat in the General Purpose Spectrometer at the Paul Scherrer Institute.

Temperatures ranged from 1.6 K to 110 K. The $\mu^+$ spin polarization was aligned anti-parallel to the muon beam's momentum, with 'forward' and 'back' detectors ('FB') along the horizontal beamline on either side of the sample, and 'up' and 'down' detectors ('UD') located above and below the sample, perpendicular to the beamline. Data were analyzed using the Musrfit software [12]. In $\mu^+$SR, the spin-polarized $\mu^+$ comes to rest at some preferred interstitial crystalline stopping site or sites. The time evolution of the spin orientation is monitored by the detection of positrons emitted by the $\mu^+$ upon decay (lifetime $\tau_\mu$ = 2.2 µs); positron emission is preferentially along the $\mu^+$ polarization direction. After several million muon decay events a time histogram yields the time-dependence of the component of the spin polarization along the detector direction. The muon depolarization is caused by the presence, and variations (spatial or temporal), of the local magnetic field at the muon stopping sites.

In Fig. 1 we show the time-dependent muon polarization in the FB direction at 110, 55, and 1.6 K in zero applied magnetic field. Three general observations can be made before any detailed analysis is carried out: (1) the depolarization is slow, indicating weak magnetism; (2) the downward curvature is characteristic of $\mu^+$ depolarization via static randomly oriented and densely packed magnetic moments (Gaussian Kubo-Toyabe, or 'GKT', depolarization [11,13]); and (3) the variation with temperature is very weak. The data in Fig. 1 were well fit to the GKT depolarization function

$$P_{ZF}(t) = (1 - f_{BG})\left\{\frac{2}{3}[1 - (\sigma_{ZF}t)^2]exp\left[-\frac{1}{2}(\sigma_{ZF}t)^2\right] + \frac{1}{3}\right\} + f_{BG} \qquad (1)$$

The parameter $\sigma_{ZF}$ is related to the second moment of the field distribution, $\Delta_{ZF}$, experienced by the muon ensemble: $\sigma_{ZF} = \gamma_\mu \Delta_{ZF}$, where $\gamma_\mu = 2\pi(135.5$ MHz/T) is the muon gyromagnetic ratio. A small correction for the fraction of muons landing outside the sample, $f_{BG}$, is also used. $f_{BG}$ was fit from the low temperature data and fixed at the value 0.071. The results for $\sigma_{ZF}$ are shown in Fig. 2 (open circles). The values are of order 0.2 MHz, not unexpected for the densely packed nuclear moments of 5.2, 3.4, and 2.5 $\mu_N$ due to the $^{51}$V, $^{121}$Sb, and $^{123}$Sb nuclei, respectively (see discussion to follow). Finally, we note that application of a weak magnetic field of 25 G along the initial muon polarization direction is sufficient to completely decouple the muon from the internal magnetic field and prevent depolarization (see the inset of Fig. 1), demonstrating that the internal field depolarizing the muon is static on the muon timescale [13]. The calculated Kubo-Toyabe depolarization in the presence of a longitudinal 25 G field and with $\sigma_{ZF}$ fixed at its low temperature zero-field value is also included in the inset, and matches the data at $T$ = 1.6 K quite well.

To more accurately probe the temperature variation of the depolarization, we applied a 100 G field transverse to both sets of detector lines, and monitored the relaxation of the muon precession due to disorder in the local magnetic field at the muon stopping sites. The data were fit to an oscillatory depolarization with a Gaussian envelope

$$P_{TF}(t) = (1 - f_{BG})\{\cos[\gamma_\mu B_{loc} t + \phi] \exp[-\tfrac{1}{2}(\sigma_{TF} t)^2]\} + f_{BG} \cos[\gamma_\mu B_{loc} t + \phi] \ . \quad (2)$$

Both FB and UD data sets were fit simultaneously to more accurately determine the parameters $\sigma_{TF}$ and $B_{loc}$; $\sigma_{TF}$ is related to the second moment of the local field distribution, and the average local field at the muon stopping site is $B_{loc}$. $\phi$ is the phase angle of the initial polarization relative to the FB detector orientation, fit to be -6° for the FB detectors, and fixed at 84° for the UD detectors. The background fraction $f_{BG}$ was again fixed at 0.071. The fit results for $\sigma_{TF}$ and $\sigma_{ZF}$ as a function of temperature are shown in Fig. 2, while the inset shows as an example the data and curve fit for the UD detectors at $T$ = 1.6 K.

The depolarization at all temperatures is dominated by densely packed quasistatic moments as manifest in the GKT (zero field) and Gaussian envelope (transverse field) depolarization fit functions. Our observed range of values for $\sigma_{TF}$ is very close to that observed at low temperatures in pure vanadium metal [14] where depolarization is due solely to the $^{51}$V nuclear moments, strongly suggesting that our observed depolarization is also due to nuclear moments. To confirm this, calculations of the nuclear contribution to zero-field depolarization, $\sigma_{ZF,Nuc}$, were performed using the semi-classical equation

$$\sigma_{ZF,nuc}^2 = 2 \left(\frac{\gamma_\mu \mu_0}{4\pi}\right)^2 \sum_{i=1}^{N_{ns}} \frac{\gamma_i^2 \hbar^2}{r_i^6} \frac{I_i(I_i + 1)}{3} \quad (3)$$

as given in Ref. 11. The sum was taken over all nuclei within 20 Å of the muon stopping site; $I_i$, $\gamma_i$, and $r_i$ are the spin, gyromagnetic ratio, and distance from the muon site of each nucleus. To account for the two isotopes of antimony, $^{121}$Sb and $^{123}$Sb, which have different nuclear moments, the calculation was performed 100 times and averaged, with each antimony nucleus being randomly assigned as $^{121}$Sb or $^{123}$Sb each time. Scanning various muon stopping sites throughout the unit cell we find a range of depolarization rates 0.13 $\mu s^{-1}$ < $\sigma_{ZF,nuc}$ < 0.56 $\mu s^{-1}$, consistent with our observations.

High temperature magnetic susceptibility data presented in Ref. 7 were consistent with either one-third of the vanadium ions carrying a small 0.22 $\mu_B$ magnetic moment, or dilute spin-1/2 impurities at a concentration of 1.5%. For the former case we have again calculated the expected depolarization rate, now due to weak electronic moments, for various muon sites throughout the unit cell, and find that depolarization rates would be in the range 2.5 $\mu s^{-1}$ < $\sigma_{TF,elec}$ < 25 $\mu s^{-1}$. Our results are at least an order of magnitude smaller than this, and we rule out this scenario. In the latter case the muon depolarization would be described by an exponential depolarization function $P_i = exp(-\lambda t)$ with an estimated rate of $\lambda$ = 0.25 $\mu s^{-1}$. This functional dependence is clearly inconsistent with our observed time-dependent depolarization, and so this possibility is also ruled out.

We now consider spin clusters consisting of magnetic vanadium (V$^{4+}$) ions, which were proposed as the origin of the UAHE in Ref. 8. The clusters would have short-ranged correlations, but no long-range ordering. While the Gaussian Kubo-Toyabe description is not applicable, the Gaussian-broadened Gaussian [11, 15] is a variation on the GKT accounting for correlations, and would yield similar rates and qualitative shapes as described above for the disordered 0.22 $\mu_B$ moments when averaging the muon response over a macroscopic volume of sample (see for example results for $Na_4Ir_3O_8$ [16]). Our results show no evidence for such short-range correlations.

We conclude that depolarization is dominated by the nuclear moments present in the system, with no significant contribution from electronic local moments. The modest temperature dependence of $\sigma_{ZF}$ and $\sigma_{TF}$ tracks the bulk susceptibility [7] – including a slight decrease cooling through 80 K, and subsequent increase upon further cooling below 50 K - and this rules out any muon-specific origins for the temperature dependence, e.g. muon diffusion or local lattice distortions due to the charge of the implanted muon. At present we do not offer any specific explanation for the temperature dependence, as such a small change in depolarization could be caused by any number of subtle intrinsic mechanisms that couple to the bulk susceptibility.

Are there scenarios for which the local $V^{4+}$ moments exist but do not depolarize the muons? It is possible that the moments fluctuate on timescales less than a nanosecond, essentially becoming invisible to the muons, but such high frequency fluctuations seem unlikely for single ions at low temperatures, and less so for three-spin clusters. Recently, it was theoretically shown that muons would be insensitive to chiral paramagnetic correlations [17]. Given the theoretical work showing the possibility of a chiral spin liquid hosted on a kagome lattice [18, 19] this presents an interesting possibility for further investigation.

As an alternative explanation, electron delocalization in the system could invalidate the ionic limit electron counting arguments presented in Ref. 7, and there are no $V^{4+}$ ions. This is not unreasonable given that the V-V separation of 2.74 Å in $KV_3Sb_5$ [7] is comparable to the 2.62 Å separation for weakly paramagnetic metallic vanadium [20], for which muon depolarization is dominated by nuclear moments. Similarly, mixed valency of the V ions as induced by correlations would considerably weaken the magnetic response, as proposed early on for $VO_2$ [21].

A more interesting alternate scenario to account for the weak magnetism is one in which pairs of $V^{4+}$ ions dimerize via orbital ordering, forming a non-magnetic singlet state, roughly in analogy with the well-studied case of dimer formation in $VO_2$ [22]. Following this model the muon depolarization would be dominated by nuclear depolarization. This is observed for $VO_2$, where results very similar to ours are reported in the temperature range 50 – 350 K [23]. In $KV_3Sb_5$, a minority of unpaired $V^{4+}$ ions could account for the Curie-like weak paramagnetism observed below 80 K. This highly dilute collection of 'impurity' spins could add a small additional channel for depolarization of muons, accounting for the weak temperature dependence of $\sigma_{TF}$ which tracks the bulk susceptibility. While the anomalies observed near 80 K in heat capacity, susceptibility [7], and $\frac{d\rho_{xx}}{dT}$ [8] are suggestive of the opening of a spin dimer gap, no direct evidence for a structural transition has been observed [7], although such a transition is observed in $CsV_3Sb_5$ [5]. Studies directed at breaking dimerized V-V pairs, e.g., off-stoichiometry (deintercalation) effects, impurity doping and the effects of strain, would be excellent tests of the applicability of this model to $KV_3Sb_5$.

Regardless of the origin of the weak magnetism, our results suggest that additional mechanisms beyond scattering from spin clusters should be investigated in order to fully understand the remarkable UAHE reported in Ref. 8.

In conclusion, we have used the highly sensitive local probe technique $\mu^+$SR to study the kagome conductor $KV_3Sb_5$. We find no evidence of vanadium local moment formation, but rather a muon response that is dominated by the weak nuclear moments in the system. This result is inconsistent with the formation of local moment spin clusters, as postulated in Ref. 8 to explain their observation of a giant anomalous Hall effect. Our muon depolarization is observed to have a weak temperature dependence component that tracks the bulk susceptibility, including anomalies observed at 50 and 80 K. We outlined several possible scenarios that may explain the weak magnetism in this system. Additional studies, utilizing a variety of techniques, are required to fully understand the interplay between band effects and magnetism in this topological system.

**Acknowledgements**

This work is based on experiments performed at the Swiss Muon Source SµS, at the Paul Scherrer Institute, Villigen, Switzerland. S.D.W. and B.R.O. acknowledge support from the University of California Santa Barbara Quantum Foundry, funded by the National Science Foundation (NSF DMR-1906325). Research reported here also made use of shared facilities of the UCSB MRSEC (NSF DMR-1720256). B.R.O. also acknowledges support from the California NanoSystems Institute through the Elings Fellowship program.

Figure 1

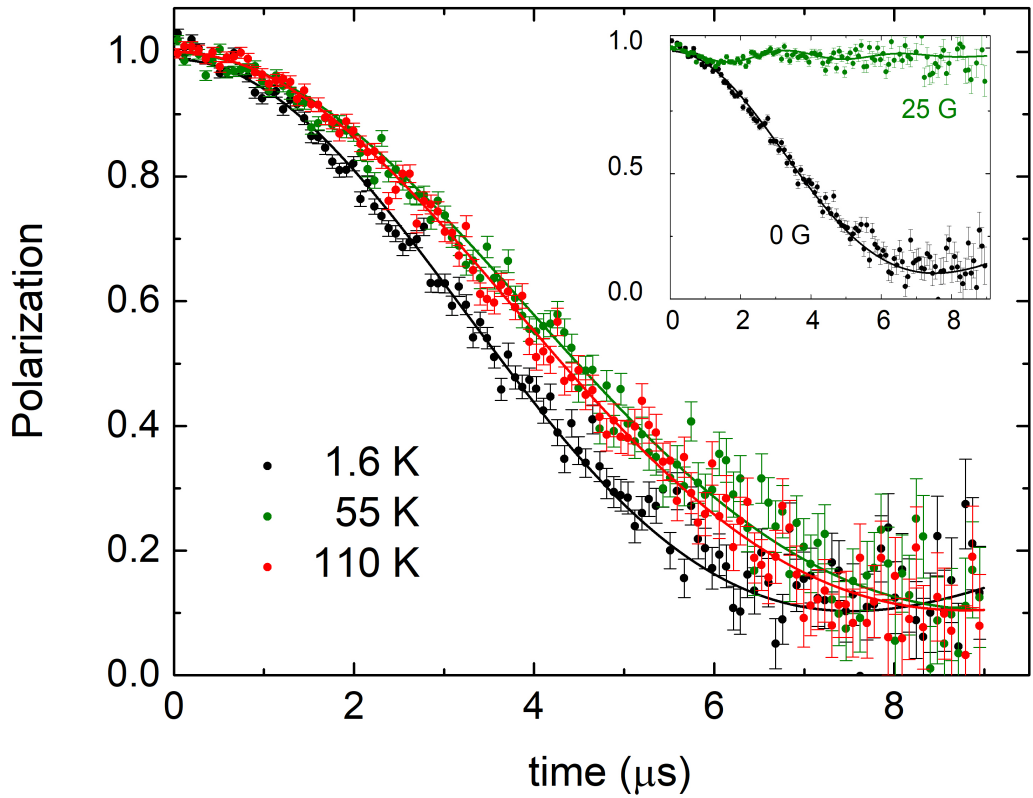

Figure 1. Zero field depolarization curves at three temperatures, along with fits as described in the text (solid lines). Inset: Data taken at 1.6 K in both zero field and in an applied longitudinal field of 25 G, along with the fit to the data at zero field, and the calculated depolarization using the zero field parameters with a 25 G longitudinal field applied (solid lines).

Figure 2

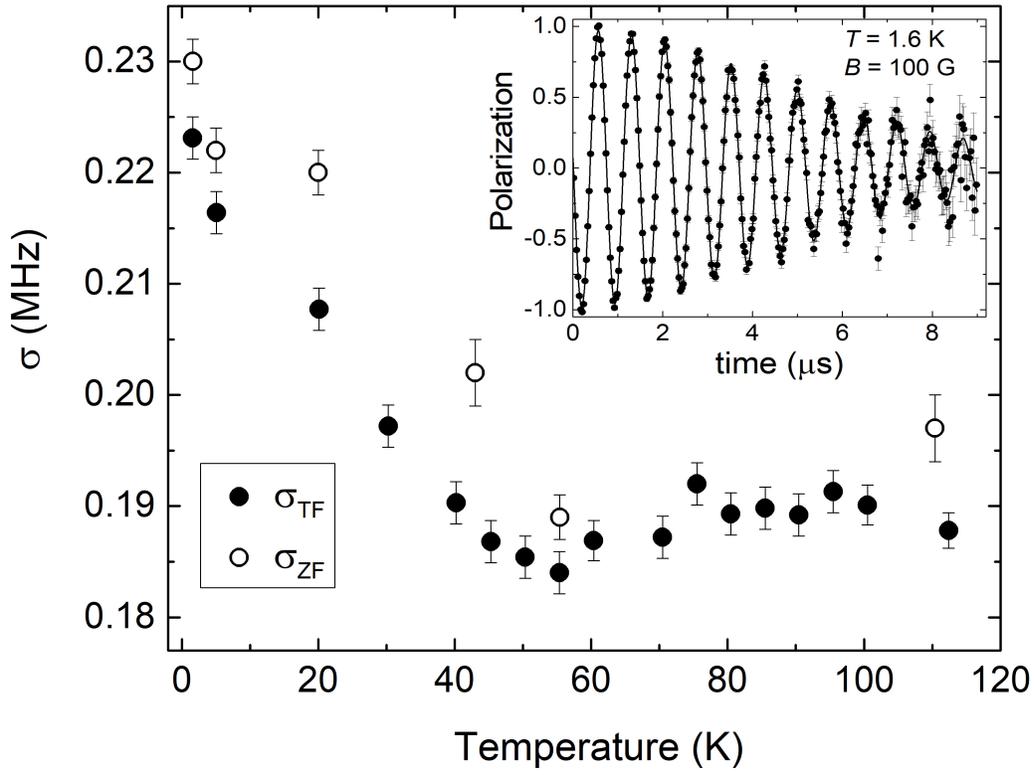

Figure 2. Temperature dependence of the Gaussian depolarization rates in zero-field (ZF) and transverse field (TF). Inset: Typical low temperature transverse field depolarization data for the UD detectors (filled circles) and the fit as described in the text (solid line).